\documentclass{ieeeaccess}
\usepackage{cite}
\usepackage{amsmath,amssymb,amsfonts}
\usepackage{algorithmic}
\usepackage{graphicx}
\usepackage{textcomp}

\usepackage{multirow}
\usepackage{comment}
\usepackage{hyperref}
 
\newcommand{\review}[1]{\textcolor{blue}{#1}} % please insert

\graphicspath{{./fig/}} %helpful if your graphic files are in another directory

\def\BibTeX{{\rm B\kern-.05em{\sc i\kern-.025em b}\kern-.08em
    T\kern-.1667em\lower.7ex\hbox{E}\kern-.125emX}}
\begin{document}
\history{Preprint - Version 1.}
\doi{n/a}
%\history{Date of publication xxxx 00, 0000, date of current version xxxx 00, 0000.}
%\doi{10.1109/ACCESS.2022.0122113}

% **************************************
%   Title & Authors
% **************************************

\title{A Unified Taxonomy for Automated Vehicles: Individual, Cooperative, Collaborative, \\On-Road, and Off-Road}
\author{\uppercase{Fredrik Warg}\authorrefmark{1},
\uppercase{Anders Thorsén}\authorrefmark{2},
\uppercase{Victoria Vu}\authorrefmark{3}
and \uppercase{Carl Bergenhem}\authorrefmark{4}
}

\address[1]{RISE Research Institutes of Sweden, Borås, Sweden (e-mail: fredrik.warg@ri.se)}
\address[2]{RISE Research Institutes of Sweden, Borås, Sweden (e-mail: anders.thorsen@ri.se)}
\address[3]{Semcon Sweden AB, Göteborg, Sweden (e-mail: victoria.vu@semcon.com)}
\address[4]{Qamcom Research and Technology AB, Göteborg, Sweden (e-mail: carl.bergenhem@qamcom.se)}
\tfootnote{This research has been supported by the Strategic vehicle research and innovation (FFI) programme in Sweden, via the project SALIENCE4CAV (ref. 2020-02946).}

\markboth
{Warg \headeretal: A Unified Taxonomy for Automated Vehicles...}
{Warg \headeretal: A Unified Taxonomy for Automated Vehicles..}

\corresp{Corresponding author: Fredrik Warg (e-mail: fredrik.warg@ri.se).}

% **************************************
%   Abstract
% **************************************

\begin{abstract}
Various types of vehicle automation is increasingly used in a variety of environments including road vehicles such as cars or automated shuttles, confined areas such as mines or harbours, or in agriculture and forestry. In many use cases, the benefits are greater if several automated vehicles (AVs) cooperate to aid each other reach their goals more efficiently, or collaborate to complete a common task. Taxonomies and definitions create a common framework that helps researchers and practitioners advance the field. However, most existing work focus on road vehicles. In this paper, we review and extend taxonomies and definitions to encompass individually acting as well as cooperative and collaborative AVs for both on-road and off-road use cases. In particular, we introduce classes of collaborative vehicles not defined in existing literature, and define levels of automation suitable for vehicles where automation applies to additional functions in addition to the driving task.
\end{abstract}


\titlepgskip=-21pt

\maketitle

% **************************************
%   Introduction
% **************************************

\section{Introduction}
\label{sec:introduction}

\PARstart{A}{utomated} vehicles (AVs) are being developed for a multitude of operational environments. While automated cars have attracted the most attention, especially in public media, automation is also attractive in domains such as freight transport, agriculture and forestry, marine operations, enclosed areas (e.g., mines, harbors and construction sites), drones, last-mile package delivery and rail-bound traffic. The aim with automation is often increased efficiency, safety, and accessibility, but also enabling new services not available without the automation \cite{ertrac2022CCAMRoadmap}.

In many applications, interaction between multiple AVs can further improve efficiency or even be essential to complete a task \cite{malik2021CollaborativeAD}. In this paper we emphasise cooperative and collaborative AVs. Cooperative AVs are multiple vehicles interacting for mutual benefit, but where each AV still has its own individual strategic goal. For road vehicles, this is often referred to as cooperative driving automation (CDA) and can be exemplified by automated cars on public roads coordinating passage through an intersection to improve traffic flow, thus reducing queuing time for all participants. We additionally define collaborative AVs as multiple AVs with a common strategic goal, collaborating to complete a joint task. Consider an automated digger loading an automated truck with gravel, where the latter proceeds to transport the material to another location. In this case, two collaborating AVs were necessary to complete the task.

For successful development of automated vehicles, creating a common understanding of the potential design space and unsolved challenges -- including the difficult issue of making sure the automation is safe -- is vital. Considerable progress has been made in establishing standard definitions and taxonomies. However, thus far, a big part of the work has been focused on on-road traffic and AVs with individual strategic goals.

In this paper, we review and discuss the existing terminology when it comes to the \textit{system-human relationship} (or levels of automation), where the most known taxonomy is the five levels of automation proposed by SAE in the standard J3016 \cite{SAE_j3016_2021}, and the \textit{system-system relationship}, which includes the CDA taxonomy defined in SAE J3216 \cite{SAE_J3216_2021}. Based on the existing work, we contribute by proposing a unified taxonomy that integrates the different types of AVs: individual, cooperative, and collaborative, and combine them with a definition of levels of automation intended to be more generally applicable than only to road vehicles. Other domains may have different conditions and challenges, e.g., additional tasks than the driving may be automated for many types of machines, and the tasks could be centrally controlled by e.g., a traffic management system (TMS).

This first preprint version of the paper is released to allow for early feedback. In future expanded revisions, we intend to expand the discussion to the implication on automation and interaction classes on terms such as operational design domain (ODD), and dynamic driving task (DDT). For the latter, we use the wider concept of dynamic operational task (DOT) which can include other tasks in addition to the driving, e.g., operating the boom and cab of a digger. We also intend to discuss possible architectural solutions, as well as typical example use cases for different classes.

The rest of the paper is organized as follows: Background on general terms for AVs, taxonomies for the human-system dimension (automation levels), and the system-system relationship (types of cooperation and collaboration) is discussed in Section \ref{sec:bg}, a new unified taxonomy is presented in Section \ref{sec:newtaxonomy}, and we conclude the work and discuss future work in Section \ref{sec:conclusions}.

%In addition to this new taxonomy, we further discuss the implication for the different types of AVs when it comes to established terms such as operational design domain (ODD), and dynamic driving task (DDT). For the latter, we use the wider concept of dynamic operational task (DOT) which can include other tasks in addition to the driving, e.g., operating the boom and cab of a digger. We also explore the design space of cooperative and collaborative vehicles, which includes how strategic and tactical decision-making is distributed, and the implications for mixed traffic environments where vehicles of different automation levels as well as unprotected users may share the same operating space.

% **************************************
%   Related Work
% **************************************
\section{Background \& Related Work}
\label{sec:bg}

This paper draws upon a large body of work within the domain of automated vehicles, as the aim is to unify and generalize concepts used for various use cases, in particular cooperative and collaborative vehicles. To that end, we begin by introducing some basic terminology needed for the rest of the paper in Section \ref{subsec:basicterm}. These terms are widely accepted, and most of them originates from the domain of individually acting on-road AVs. 
%Section \ref{sec:newconcepts} discusses how some of these concepts can be understood in the context of cooperative and collaborative vehicles. 
In Section \ref{subsec:syshum} we discuss different proposed taxonomies for classifying driving automation systems according to capability with respect to a human user, and in Section \ref{subsec:syssys} we similarly examine terminologies for the relationship between systems, either several AVs or between AV and infrastructure. Drawing upon this material, in Section \ref{sec:newtaxonomy} we propose a unified taxonomy for AVs with the system-system dimension including individual, as well as cooperative and collaborative vehicles, and the human-system dimension with definitions applicable beyond on-road vehicles.

% ----------------------------------------------
% Basic terminology
% ----------------------------------------------
\subsection{Basic Terminology for Automated Vehicles}
\label{subsec:basicterm}

Many of the well established terms in the AV domain originates from the SAE J3016 standard \cite{SAE_j3016_2021}, which has been published in several editions, the first in 2014 \cite{SAE_j3016_2014} and the latest, at the time of writing this paper, in 2021 \cite{SAE_J3216_2021}. J3016 has also been used as the base for ISO PAS 22736:2021 \cite{ISO_PAS22736}, which uses the same basic terminology. These standards define a system that has some automation pertaining to the driving task as a driving automation system, and for more advanced automation, the system is called an automated driving system (ADS). Automation levels are further discussed in Section \ref{subsec:syshum}. Automation with cooperative capabilities are defined in SAE J3216 \cite{SAE_J3216_2021} as cooperative driving automation (CDA), and an ADS with CDA capabilities is called a cooperative ADS (C-ADS). Machine interaction is further discussed in Section \ref{subsec:syssys}.

Key defining characteristics for a driving automation system are the operational design domain (ODD), which defines the operating conditions for which a specific driving automation feature is designed to handle. Over the years, several frameworks for describing these operational conditions have been proposed. At the time of writing, an international standard, ISO 34503 \cite{ISO_34503}, is close to completion. Even though not explicitly discussed in J3016, the ODD is also regarded as a key concept for safety assurance, as it defines the limits of the capability of the automated function. This is discussed e.g., by Gyllenhammar et al. \cite{oddpapper} and ISO TR 4804 \cite{ISO_TR4804} which discusses how to approach safety and security for an ADS. Another important concept is the dynamic driving task (DDT), which consists of the sub-tasks of motion control (managing steering, brake, acceleration) and object and event detection and response (OEDR), i.e. monitoring the surroundings and reacting to events external to the vehicle such as pedestrians and the actions of other vehicles. Furthermore, it is recognized that AVs may experience problems preventing them from completing their user defined strategic mission. This may happen either if the ADS experiences a performance-critical failure, or the vehicle is close to exiting its ODD. In these cases, the AV is expected to be able to come to a stable stopped state called a minimal risk condition (MRC), with the purpose to reduce the risk of a crash. Gyllenhammar et al. \cite{gyllenhammar2021minimal} refine the definition to say that it should be a position with an acceptable risk, and that this will depend on (i) the risk of the selected position, (ii) the frequency to enter MRC, and (iii) the rate of recovery (how long the vehicle remains in a position which may not be sufficiently safe in the long term). The maneuver necessary to transition from the original mission to the MRC is called the minimal risk maneuver (MRM) in ISO TR 4804 \cite{ISO_TR4804} and DDT fallback in J3016 \cite{SAE_j3016_2021}. These and some other key terms are summarized in Table \ref{tab:basic_terms}, and are important for the understanding of the rest of the paper.

%--- TABLE Basic Terms ---
\begin{table}[htbp]
\begin{center}
\caption{\textbf{Basic terminology for automated driving systems.}}
\setlength{\tabcolsep}{3pt}
\renewcommand{\arraystretch}{1}
\begin{tabular}{|p{2.7cm}|p{5.3cm}|} % 8 cm 
\hline
Term & Description \\ \hline\hline
Level of driving automation & A classification of the capabilities of a driving automation system with respect to a human user. The most common taxonomy is the one defined in SAE J3016 (see subsection \ref{subsec:syshum}). \\ \hline
Driving automation system (DAS) & Automated system that performs all or part of the driving task (see also definition of DDT) on a sustained basis. \\ \hline
Driving automation feature & A feature that is a specific automated task implemented by a DAS, e.g., highway pilot or valet parking. A feature is defined by the use case, automation level, and ODD. \\ \hline
Automated driving system (ADS) & A driving automation system of level 3-5 according to the SAE levels of driving automation (see Table \ref{tab:sae_automationlevels}). \\ \hline
Automated vehicle (AV) & A vehicle equipped with at least one driving automation feature. \\ \hline
Connected and automated vehicle (CAV) & Combination of driving automation and functionality depending on V2X. \\ \hline
Cooperative driving automation (CDA) & Automation that uses communication to enable cooperation among two or more entities. \\ \hline
Cooperative-ADS (C-ADS) & An ADS with cooperative capabilities. \\ \hline
%Connected, cooperative, and automated mobility (CCAM) & \review{is this needed?} \\ \hline
Object and Event Detection and Response (OEDR) & The function of monitoring the driving environment to detect objects and events and executing appropriate responses. (SAE J3016) \\ \hline
Dynamic driving task (DDT) & The act of operating the vehicle on an operational and tactical level, including motion control (steering, accelerating, braking), and OEDR. Different parts of the DDT can be handled by either a human user or a DAS, as defined by the automation levels. (SAE J3016) \\ \hline
Operational Design Domain (ODD) & Operating conditions under which a given driving automation system, or feature thereof, is specifically designed to function. The ODD includes, but is not limited to, the environmental, geographical, and time-of-day restrictions, and/or the requisite presence or absence of certain traffic or roadway characteristics. (SAE J3016) \\ \hline
Minimal risk maneuver (MRM), minimal risk condition (MRC) & Response given a system failure or upon an ODD exit. The response can be to perform an MRM and reach an MRC, or a user taking over the DDT. In J3016, MRM is called DDT fallback. (J3016 and ISO TR 4804). \\ \hline
Vehicle-to-X (V2X) & Used to describe various use cases of connectivity in the context of CAVs, where 'x' denotes 'everything' collectively for vehicle-to-vehicle (V2V), vehicle-to-infrastructure (V2I), vehicle-to-network (V2N), vehicle-to-device (V2D) or vehicle-to-pedestrian (V2P). \\ \hline
\end{tabular}
\label{tab:basic_terms}
\end{center}
\end{table}

%\review{There is a number of good examples included, but I am thinking if same fundamental example could be used to exemplify all different interaction levels?  Could this be pedagogic to see explain the differences? Maybe we could find a few of them like (1) the intersection and (2) the digger and truck?}

% ----------------------------------------------
% System-human
% ----------------------------------------------
\subsection{Human-System Relationship}
\label{subsec:syshum}

\begin{figure}
    \centering
    \includegraphics[width=0.5\textwidth]{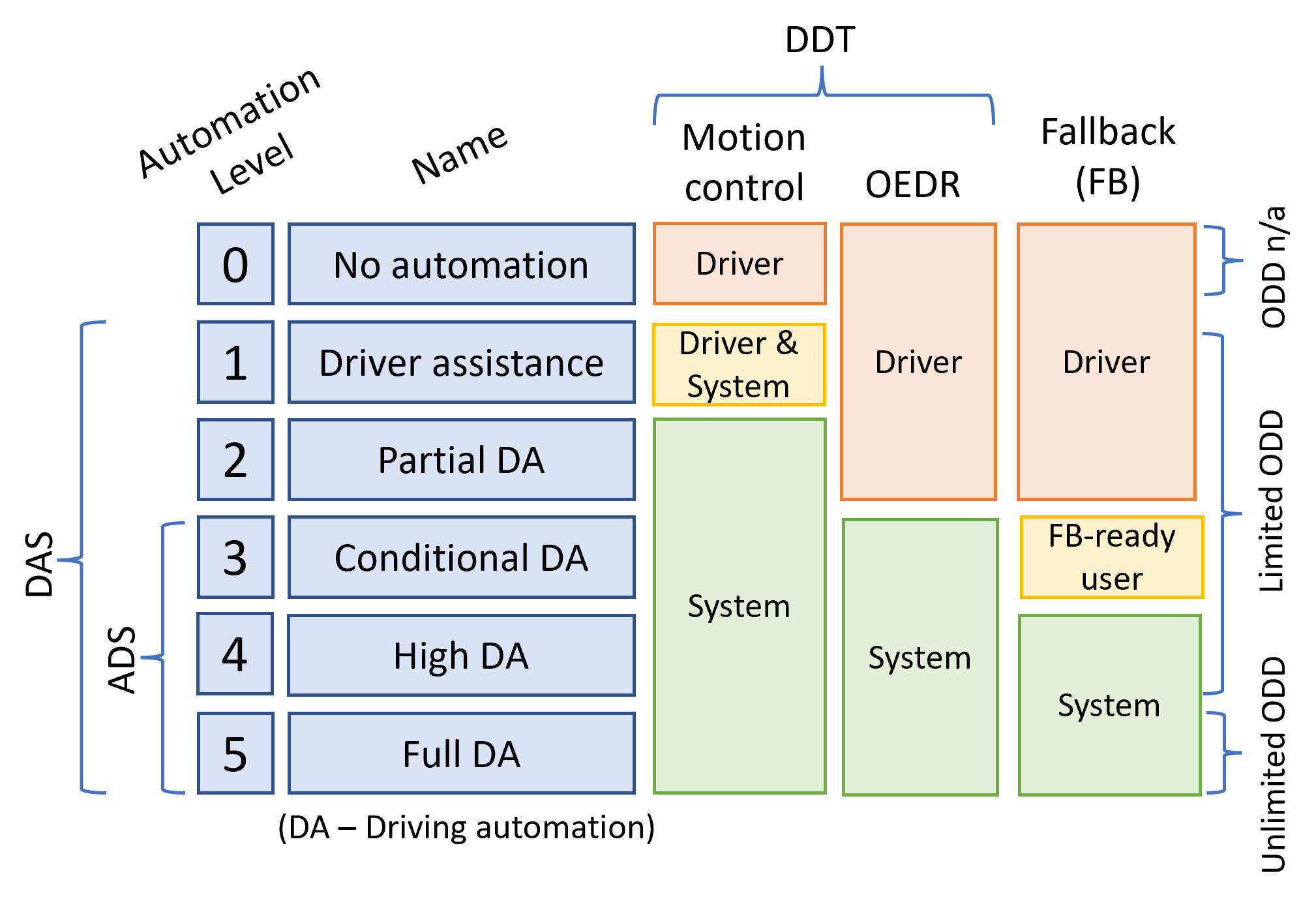}
    \caption{SAE automation levels (from \cite{SAE_j3016_2021}).}
    \label{fig:sae_levels}
\end{figure}

The primary way of describing and classifying vehicle automation has been by focusing on the division of responsibility between human users and the driving automation system using different levels or dimensions to create classes of vehicles with different automation characteristics. The most known classification is arguably the SAE levels of driving automation first defined in the standard SAE J3016:2014 \cite{SAE_j3016_2014} and still used in the latest revision J3016:2021 \cite{SAE_j3016_2021}. This standard defines automation levels 0 (no automation) to 5 (full driving automation) as described in Table \ref{tab:sae_automationlevels}. Figure \ref{fig:sae_levels} visually illustrates the difference between the levels in terms of the user control and the ODD. While the SAE taxonomy has been useful when discussing AVs, it has also been criticised for ambiguities and misuse, partly depending on having an engineering-centric view rather than a user-centric view \cite{koopman_AV_taxonomy, mobileye_AV_taxonomy} and other drawbacks \cite{mobileye_black_swans} making it less useful when it comes to describing the capabilities of actual products.

%--- TABLE SAE automation levels ---
\begin{table}[htbp]
\begin{center}
\caption{\textbf{SAE levels of driving automation (from \cite{SAE_J3216_2021}).}}
\setlength{\tabcolsep}{3pt}
\renewcommand{\arraystretch}{1}
\begin{tabular}{|p{0.7cm}|p{2cm}|p{5.3cm}|} % 8 cm 
\hline
Level & Name & Description \\ \hline\hline
0 & No automation & Driver performs the entire DDT. Driver can be aided by active safety systems. \\ \hline
1 & Driver assistance & System performs sustained lateral or longitudinal control within a limited ODD. Driver performs the rest of the DDT. \\ \hline
2 & Partial driving automation & System performs sustained lateral and longitudinal control within a limited ODD. Driver performs the rest of the DDT and supervises the system. \\ \hline
3 & Conditional driving automation & Systems performs the entire DDT within a limited ODD, but a fallback-ready user is expected to be receptive to requests for intervention. \\ \hline
4 & High driving automation & System performs the DDT, and DDT fallback if needed, within a limited ODD. No user intervention necessary. \\ \hline
5 & Full driving \mbox{automation} & System performs the DDT and DDT fallback with an unlimited ODD (i.e., has at least the same capabilities as a proficient human driver for all on-road conditions). No user intervention necessary.\\ \hline
\end{tabular}
\label{tab:sae_automationlevels}
\end{center}
\end{table}

Attempts to introduce different definitions have been made to stem the confusion, however, at the time of writing, none of these are yet as widely accepted as the SAE levels. Koopman \cite{koopman_AV_taxonomy} proposes four operational modes according to Table \ref{tab:koopman_classes}. In his view, the category supervised automation should be rather limited, at least considering today's technology. Taking into account the capabilities of ordinary drivers to act as a back-up, it would limit the use to lane keeping and speed control on highways, as manoeuvres such as turning at an intersection would be too dangerous. He also specifically includes a category for test purposes, where the human user is a trained safety driver expected to be more able than the average driver to handle automation failures. The proposal from AV technology provider Mobileye has similarities to Koopman's proposal, but is based on four dimensions: hands-on/hands-off (steering wheel), eyes-on/eyes-off (the road), driver/no driver, and MRM requirement. Using these dimensions, four vehicle categories, shown in Table \ref{tab:mobileye_classes}, are defined. Compared to the other taxonomies, this proposal also discusses the role of human drivers (in-vehicle or teleoperators) for what they term non-safety-related situations to resolve traffic situations after the AV has come to a stopped state. E.g., if the AV would stop due to a policeman sign, a human driver might be necessary to resume the trip to prevent the vehicle from remaining in a position where it blocks other traffic. An even simpler classification is used by ADAS and AD software company Zenseact, which differentiates only between supervised and unsupervised functions to make clear what the expected role of the human user is \cite{zenseact_whitepaper}.

%\review{Different users: remote driver, remote assistance, user, etc. Include figure or in terminology table? Where to discuss?}

%--- TABLE Koopman automation levels ---
\begin{table}[htbp]
\begin{center}
\caption{\textbf{Automation categories proposed by Koopman (from \cite{koopman_AV_taxonomy}).}}
\setlength{\tabcolsep}{3pt}
\renewcommand{\arraystretch}{1}
\begin{tabular}{|p{2.7cm}|p{5.3cm}|} % 8 cm 
\hline
Category & Description \\ \hline\hline
Driver assistance & Licensed human driver performs driving task. Vehicle systems provide active safety features, driver support, and convenience functions. \\ \hline
Supervised \mbox{Automation} & Licensed human driver monitors road and intervenes for situations the vehicle is not designed to handle, including tasks beyond ordinary lane-keeping. Vehicle provides lane-keeping and speed control. \\ \hline
Autonomous \mbox{Operation} & No human driver. Steering wheel optional. Vehicle is responsible for driving and safety. \\ \hline
Vehicle Testing & Trained safety driver mitigates dangerous behaviour and at times performs the driving. Vehicle feature under test is expected to exhibit dangerous behaviours. \\ \hline
\end{tabular}
\label{tab:koopman_classes}
\end{center}
\end{table}

%--- TABLE Mobileye automation dimensions ---
\begin{table}[htbp]
\begin{center}
\caption{\textbf{Automation categories proposed by Mobileye (from \cite{mobileye_AV_taxonomy}).}}
\setlength{\tabcolsep}{3pt}
\renewcommand{\arraystretch}{1}
\begin{tabular}{|p{2.7cm}|p{5.3cm}|} % 8 cm 
\hline
Category & Description \\ \hline\hline
Eyes-on/hands-on & Driver assistance functions. Driver still responsible for full driving task. \\ \hline
Eyes-on/hands-off & Driver supervises the system but the system takes control of the driving in a limited ODD. Driver monitoring can be used to increase safety. \\ \hline
Eyes-off/hands-off & The system performs the driving within a limited ODD without human supervision. The system must be able to conduct an MRM if the driver does not resume the driving task upon ODD exit. Human driver present is required for control in non-safety-related situations. \\ \hline
No driver & No human driver present. Teleoperator may be used to resolve non-safety-related situations. \\ \hline
\end{tabular}
\label{tab:mobileye_classes}
\end{center}
\end{table}

All of these automation level definitions are focused on on-road traffic, and the driving task. In Section \ref{sec:newtaxonomy} we propose a similar classification, but include other automation tasks making the classification suitable also for vehicles with other forms of automation, which is applicable for many machines.

% ----------------------------------------------
% System-system
% ----------------------------------------------
\subsection{System-System Relationship}
\label{subsec:syssys}

Another dimension in the design space for AVs is interaction between agents in a diverse environment, e.g., between several AVs, between AVs and infrastructure, or between AVs and non-automated traffic participants, i.e., vehicles or pedestrians that have a capability to communicate with AVs but are not themselves controlled by automation. 

There are numerous works published related to interaction between AVs, often called cooperative or collaborative AVs. Much of the presented works are from the intelligent transport system (ITS) domain as it relies on V2X communication and is often combined with digital twins, see, e.g., \cite{hui2022CollaborationServiceDigitalTwinEnabled,zhang2022CollaborativeDrivingLearningAided, wang2022DigitalTwinAssistedCooperative, slamnik-krijestorac2021Collaborativeorchestrationmultidomain}. Cooperative and collaborative driving is also in the scope of System-of-systems research, see, e.g, \cite{pelliccione2016, pelliccione2020connectedcarssystems}. Much of the work uses either the terms cooperative or collaborative for all types of AV interaction, or uses both terms interchangeably. Later, in Section \ref{sec:newtaxonomy}, we make an attempt to clearly define the difference between these two types of AVs, as we believe there is a significant difference when it comes to the strategic goal which should be reflected in the taxonomy. While we use these terms consistently with our own definition, it is important to notice that the terms may be used differently and inconsistently in the referenced literature.

One classification of cooperative automated vehicles has been established by SAE in the standard J3216 \cite{SAE_J3216_2021}. This classification defines four different types of cooperation between on-road vehicles, shown in Table \ref{tab:cda_classes}. We use the term cooperative (mostly) consistent with this taxonomy, which includes different strategies where several AVs communicate for mutual benefit (except possibly for the category 'prescriptive' where the action may be for the benefit of others, e.g., a road operator temporarily prescribing a re-routing of traffic for some reason). The division in classes in this taxonomy is thus based on different mechanisms or capabilities of the cooperative action.

%--- TABLE SAE CDA classes ---
\begin{table}[htbp]
\begin{center}
\caption{\textbf{SAE Cooperative Driving Automation (CDA) classes (from \cite{SAE_j3016_2021}).}}
\setlength{\tabcolsep}{3pt}
\renewcommand{\arraystretch}{1}
\begin{tabular}{|p{0.7cm}|p{2cm}|p{5.3cm}|} % 8 cm 
\hline
Class & Class name & Description \\ \hline\hline
A & Status-sharing & Sharing information, e.g. position, sensor data, or world model, to help other actors make their decisions. \\ \hline
B & Intent-sharing & Sharing intent for future actions (operational, tactical, strategical) to help other actors make their decisions. \\ \hline
C & Agreement-seeking & Communicates to reach agreement with other actors to optimise some parameter(s) for mutual benefit, e.g. non-interference or improved traffic flow. \\ \hline
D & Prescriptive & Generally individual action but accepts some temporary prescriptive actions to achieve tactical goal defined by other actor, e.g. road operator. \\ \hline
\end{tabular}
\label{tab:cda_classes}
\end{center}
\end{table}

Malik et al. present in \cite{malik2021CollaborativeAD} a survey related to cooperative/collaborative automated driving. A taxonomy is proposed covering: (a) background categories comprised of objectives, motivations, collaboration types, collaboration scopes, as well as applications, and (b) architecture categories consisting of interaction types, core components, architectural layers, manoeuvres, communication technologies, and coordination strategies. Further, aspects of cooperative driving are discussed including: (1) why collaborative driving is important to get full advantage of AD vehicles, (2) the motivations for vehicular agents to collaborate, (3) possible control strategies for collaborative vehicles divided as \textit{centralised coordination}, \textit{decentralised with coordination}, and \textit{decentralised without coordination}, (4) the potential benefits of collaborative driving divided into the categories traffic efficiency, safety and miscellaneous, and (5) the collaboration types described in Table \ref{tab:malik_types}. It should be noted that Malik et al. mixes what we in this paper understand as the distinct types of cooperative and collaborative interaction. Unlike J3216, the cooperation/collaboration types of Malik et al. are not mainly based on the capabilities or the cooperative or collaborative action, but rather whether it is obligatory or voluntary for the individual AVs to participate.

%--- TABLE Malik Collaborative types ---
\begin{table}[htbp]
\begin{center}
\caption{\textbf{Collaboration types defined in \cite{malik2021CollaborativeAD}.}}
\setlength{\tabcolsep}{3pt}
\renewcommand{\arraystretch}{1}
\begin{tabular}{|p{2.7cm}|p{5.3cm}|} % 8 cm 
\hline
Type & Description \\ \hline\hline
Imperative \mbox{collaboration} & Obligatory for  to participate in collaboration. \\ \hline
Voluntary \mbox{collaboration} & Voluntary participation, depending on use case. \\ \hline
Hybrid \mbox{collaboration} & Imperative collaboration during certain period, but agents can voluntarily end collaboration. \\ \hline
\end{tabular}
\label{tab:malik_types}
\end{center}
\end{table}

One useful definition of a system-of-systems is \textit{''an assemblage if components which individually may be regarded as systems, and which possesses two additional properties: operational independence of the components [...] managerial independence of the components [...]"} \cite{maier1998Architectingprinciplessystemsofsystems}. In  system-of-systems research, the categories in Table \ref{tab:sos_types} are often used. The classification in this domain thus has a management perspective, like Malik et al. considering the degree of obligatory vs voluntary participation, but also ownership/management of the individual constituents. It can be noted that one of the categories is called collaborative system-of-system. However this definition is not consistent with the one we introduce in this paper. As the use of these terms is already inconsistent in literature, achieving consistency with all existing taxonomies simultaneously is not possible. A term we adopt from the system-os-systems domain, however, is the designation \textit{constituent} \cite{dahmann2008UnderstandingCurrentState} to refer to the individual systems within a system-of-systems (or, with the terms we commonly use later in this paper, in a cooperative or collaborative system).

%--- TABLE SoS types ---
\begin{table}[htbp]
\begin{center}
\caption{\textbf{System-of-system types (see \cite{maier1998Architectingprinciplessystemsofsystems, dahmann2008UnderstandingCurrentState}).}}
\setlength{\tabcolsep}{3pt}
\renewcommand{\arraystretch}{1}
\begin{tabular}{|p{2.7cm}|p{5.3cm}|} % 8 cm 
\hline
Type & Description \\ \hline\hline
Directed & Integrated SoS built to fulfil a specific purpose and centrally managed. Components maintain an ability to operate independently but the normal operation mode is subordinate to the central system. \\ \hline
Virtual & No central management or centrally agreed-upon purpose for the SoS. \\ \hline
Collaborative & Central management organisation without coercive power. Components must collaborate voluntarily to fulfil agreed upon central purposes. \\ \hline
Acknowledged & Recognised objectives and a designated manager exists, but the components retain independent ownership, objectives, funding, development, and sustainment. \\ \hline
\end{tabular}
\label{tab:sos_types}
\end{center}
\end{table}

In addition to classification of the constituent vehicles, there is a classification of different levels of infrastructure support for use by CAVs. These so called infrastructure suppoer for automated driving (ISAD) levels \cite{carreras2018eu} are from the EU project Inframix \footnote{\url{https://www.inframix.eu/}}. We also consider this aspect to some degree in our proposed taxonomy.

%--- TABLE ISAD levels ---
\begin{table}[htbp]
\begin{center}
\caption{\textbf{Infrastructure Support for AD (ISAD) levels (from \cite{carreras2018eu}).}}
\setlength{\tabcolsep}{3pt}
\renewcommand{\arraystretch}{1}
\begin{tabular}{|p{0.7cm}|p{2cm}|p{5.3cm}|} % 8 cm 
\hline
Level & Name & Description \\ \hline\hline
E & Conventional infrastructure & No digital information. AVs must recognise static infrastructure and dynamic events. \\ \hline
D & Static digital information & Digital map including static road signs available. AVs must recognise temporary events including traffic lights. \\ \hline
C & Dynamic digital information & All dynamic and static infrastructure information provided in digital form. \\ \hline
B & Cooperative perception & Infrastructure provides perception of microscopical traffic situations. \\ \hline
A & Cooperative driving & Infrastructure can guide AVs based on real-time information of vehicle movements to optimise traffic flow. \\ \hline
\end{tabular}
\label{tab:isad_levels}
\end{center}
\end{table}

\begin{comment}
\subsection{Architecture of Cooperative and Collaborative systems}
\label{subsec:bg_arch_cc}

In \cite{wang2022ReviewIntelligentConnected}, Wang et al. discuss the research of intelligent connected vehicle cooperative driving systems including vehicles, infrastructure, and test sites. The paper covers the subjects (1) cooperative control, including vertical formation, collaborative decision making, and collaborative positioning, (2) vehicle communication,including communication security and control strategy for communication delay, and (3) test and evaluation, including real vehicle road test platform, virtual test platform, and test method and evaluation.
\end{comment}

% **************************************
%   Considerations for Coop and Coll
% **************************************

\begin{comment}
\section{Concepts for Cooperative \& Collaborative Automated Vehicles}
\label{sec:newconcepts}

\subsection{TODO}

\subsubsection{Dynamic Operational Task}
TODO

\subsubsection{Remote Operator}
TODO

\subsubsection{Operational Design Domains}
TODO

\subsubsection{Minimal Risk Manoeuvres}
TODO \review{Note as future work to extend this, will be MRM paper}.
\end{comment}

% **************************************
%   Taxonomy
% **************************************
\section{Taxonomy for Automated Vehicles}
\label{sec:newtaxonomy}

\subsection{System-System Dimension Revisited}

In this section we propose a unified taxonomy aimed to encompass all types of AVs. The full taxonomy is shown in Table \ref{tab:newtaxonomy}. We divide AVs in three major types. \textit{Individual AVs} operate of their own, without assistance or communication with other vehicles, and pursue their own strategic goal. For \textit{cooperative AVs}, we use the already existing definitions from SAE J3216 \cite{SAE_J3216_2021}, with the additional definition that cooperative AVs are vehicles that cooperate for mutual benefit, but which have their own individual strategic goals. There are further use cases, partly covered by system-of-systems taxonomy but previously not well defined in the context of vehicle automation, where we have added the AV type denoted as \textit{collaborative AVs}. The defining difference between cooperative and collaborative AVs is that for the latter, the constituents pursue a common strategic goal. This definition covers many use cases in confined areas or fleet operations. While the words cooperative and collaborative are sometimes used interchangeably, the terms are also used in other contexts with the distinction between having a shared goal versus interacting for mutual benefit but having individual goals\footnote{See, e.g., \url{https://blog.jostle.me/blog/collaboration-vs-cooperation} and \url{https://resourced.prometheanworld.com/collaborative-cooperative-learning/}.}.

These types of AVs, and the distinct classes defined within each type, are further described and exemplified in the following subsections.

% --- TABLE Taxonomy ---
\begin{table*}[htbp]
\begin{center}
\caption{\textbf{Taxonomy of interaction for automated vehicles.}}
\setlength{\tabcolsep}{3pt}
\renewcommand{\arraystretch}{1.2}
\begin{tabular}{|p{3cm}|p{2.5cm}|p{9.5cm}|p{2cm}|} % 17 cm
\hline
Type & Class & Characteristics & Designation  \\ \hline\hline

Individual & Ego-sensing & Relies exclusively on on-board sensors and decision-making capabilities. This can include static digital infrastructure information. & IndEgo \\ \cline{2-4}

Single AV with individual strategic goal. & Connected & Uses connected services (e.g. cloud services, dynamic digital infrastructure information or infrastructure perception) to enhance sensing and/or decision-making. & IndCon \\ \hline

Cooperative  & Status-sharing & Sharing information, e.g. position, sensor data, or world model, to help other actors make their decisions (J3216 Class A). & CoopSs\\ \cline{2-4}
& Intent-sharing & Sharing intent for future actions (operational, tactical, strategical) to help other actors make their decisions (J3216 Class B). & CoopIs \\ \cline{2-4}
Multiple AVs with individual strategic goals. & Agreement-seeking & Communicates to reach (voluntary) agreement with other actors to optimise some parameter(s) for mutual benefit, e.g. non-interference or improved traffic flow (J3216 Class C). & CoopAs \\ \cline{2-4}
& Prescriptive & Generally individual action but accepts some temporary prescriptive actions to achieve tactical goal defined by other actor, e.g. road operator (J3216 Class D). & CoopPr \\ \hline

Collaborative  & Coordinated & Communicates with other constituents to reach agreements for how to act in order to achieve a common strategic goal. & CollCo\\ \cline{2-4}
& Choreographed & Multiple constituents acting individually but being designed to follow a common global scenario/goal, i.e., unlike coordinated vehicles they do not rely on communication to perform the collaborative task. & CollCh \\ \cline{2-4}
Multiple AVs with common strategic goal. & Orchestrated & Constituents are directed by a single entity acting to achieve a strategic goal. The directing entity can be one of the constituents or a separate system referred to as e.g., traffic management system (TMS). & CollOr \\ \hline

\end{tabular}
\label{tab:newtaxonomy}
\end{center}
\end{table*}

\subsubsection{Individual Classes}

For individual AVs, two distinct classes are defined. The first class is called \textit{ego-sensing} and is an AV relying exclusively on its on-board sensors for the automated task. An example can be a highway autopilot function where the AV does not need any information beyond what is provided by its own sensors such as cameras, lidar, or radar.

The second individual class is \textbf{connected} vehicles, often denoted as CAV, where the individual AV additionally may make use of information from cloud services and connected infrastructure, or even sensors that are part of the infrastructure, but not directly communicating with nearby vehicles.

\subsubsection{Cooperative Classes}

Cooperative AVs  are able to share information and coordinate their actions, but do not rely on each other to achieve their own goals. Each vehicle can operate independently, but can also benefit from information and actions of other vehicles in the system. There are four different classes of cooperation that cooperative AVs can take, status-sharing, intent-sharing, agreement-seeking and prescriptive. 

Status-sharing involves the sharing of perception data for the potential use by receiving vehicles \cite{SAE_J3216_2021}. Vehicle A sending camera information informing vehicle B of a pedestrian’s location outside its’ field of view would allow for vehicle B to use this information and move forward cautiously, now aware of an occluded road user. 

With intent-sharing, information about a vehicle’s planned actions can be shared with another vehicle, once again allowing the receiving vehicle to plan/act accordingly \cite{SAE_J3216_2021}. For example, after approaching an intersection, vehicle A, the lead vehicle, broadcasts its intent to turn left and the following vehicle, vehicle B, receives this intent, adjusting its speed and trajectory accordingly to avoid a collision. 

Agreement-seeking involves a number of messages being sent among cooperative vehicles with the intention of planning actions\cite{SAE_J3216_2021}. For example, when two vehicles approach a narrow bridge from opposite directions, the vehicles exchange information (e.g., speed, trajectory, intended actions), which is then used to negotiate a mutually agreed upon plan, ensuring that both vehicles cross the bridge safely and efficiently.

Prescriptive cooperation involves a road operator (e.g., a transportation authority) providing instructions to specific traffic participants, who then perform the specified command without need for agreement \cite{SAE_J3216_2021}. An example related to incident scene management could involve emergency vehicles communicating temporary road closures, requiring adjustments from nearby AVs. It should be noted that SAE J3216 includes fleet operations in this category. With our taxonomy including collaborative classes, most use cases for fleet operations would fit better in the category \textit{collaborative-orchestrated} described in the next section.

\subsubsection{Collaborative Classes}

While cooperative vehicles work together in order to help each other achieve their own individual goals, collaborative vehicles – in addition to having their own strategic goals – also work towards achieving a shared goal, known as the common strategic goal. This typically requires high levels of coordination among all constituents. Table \ref{tab:newtaxonomy} shows 3 types of architectural patterns which fall under the category of collaborative AVS – coordinated, choreographed and orchestrated.  

Coordinated collaborative AVs involve vehicles or machines engaged in specific tasks and communicating with each other during these tasks to achieve the common strategic goal. Consider a simple system that consists of two machines, a loader and a truck. The loader’s individual strategic goal is to remove material from its current location (location A) and deposit this load into the truck stationed next to it. As for the truck, it must remain stationary until a certain weight load has been met. Once this happens, the truck signals to the loader that it can take no further weight and the loader halts. The truck drives to another location (location B) where the material is then deposited. The truck then returns to location A, signals it’s return to the digger, and the cycle continues. Both the loader and truck have their own strategic goals (i.e., digging up/depositing material and moving the material), though they must work in tandem – communicating in a coordinated manner– to achieve the shared goal of removing and relocating the material from one area to another, as neither machine is able to independently accomplish this goal on its own.

Choreographed collaborative AVs refer to a group of vehicles that work together in a predetermined sequence, with each vehicle having its own defined role, rather than communicating to coordinate their actions. Together, they are able to achieve a shared goal. If the digger and truck in the example above did not use communication to signal each other but the system was rather designed so that each vehicle has its own way of determining when to proceed, the system would be choreographed. E.g., the truck could be designed to automatically leave when loaded above a certain weight, while the digger could be designed to continue filling trucks as long as they are parked in the digger area of operation.

%a platoon of three trucks travelling on a highway have been programmed to maintain a specific formation, ensuring consistent speed and improved fuel efficiency.

Orchestrated collaborative AVs involve multiple machines or vehicles that are managed and directed by a central entity to achieve a common strategic goal. In this scenario, each individual vehicle has limited autonomy and the central authority has complete control over all constituents' actions. An example of this could be a fleet of autonomous taxis that are directed by a central authority ensuring route optimisation, safety and reduced emissions. 

%To the best of the authors' knowledge, the definition of collaborative AVs in this paper is novel. Rather, previous works have used the terms cooperative and collaborative synonymously when referring to AVs that work together, regardless of potential differences \cite{malik2021CollaborativeAD}. This paper has provided a distinction between the terms cooperative and collaborative in reference to AVs, with the belief that this information can be used to enhance understanding and provide appropriate measures(?)(in this case MRMs) to improve safety.

It should be noted that some functions may be classified differently depending on application and specific implementation. For instance, platooning could be classified as collaborative if consisting of a fleet with the same destination collaborating with a joint task, or cooperative if the platoon is constituted temporarily on a highway for mutual benefit, but the objectives and destinations of the constituents differ.

%\review{Note2: Some functions may be cooperative or collaborative depending on application. Platooning can be collaborative if consisting of a fleet with the same destination. Platooning of vehicles with individual destinations is a border-case, would call it cooperative but another viewpoint would be that they are collaborative during a certain period of time, compare to Malik 'hybrid collaboration'}

\subsection{Human-System Dimension Revisited}

%\review{Remote driver and remote assistance? How do they fit in table? See J3016 and Mobileye taxonomy. I think remote operation would be manual or (more likely) assisted, while remote assistance would be supervised. If a remote user takes over control for "non-safety-related situations," this would constitute a handover and a mode switch in the automation level.}
%If a remote user goes in for "non-safety-related situations" that would be a handover and mode switch in automation level

In this section, we define a taxonomy for the human-system dimension, or classification of level of automation with respect to the involvement of a human user. We use a classification similar to Koopman \cite{koopman_AV_taxonomy}, with the main difference that the automation is not meant to include only the driving task, as many types of machines have other aspects of the operation automated, which is integral to its use and also safety-critical. As future work, we intend to further investigate this aspect. Our updated definitions are presented in Table \ref{tab:av_automation}.

%--- TABLE Automation levels ---
\begin{table}[htbp]
\begin{center}
\caption{\textbf{Levels of vehicle automation - Adapted for general AVs.}}
\setlength{\tabcolsep}{3pt}
\renewcommand{\arraystretch}{1}
\begin{tabular}{|p{2.7cm}|p{5.3cm}|} % 8 cm 
\hline
Category & Description \\ \hline\hline
Manual operation & Human user (e.g., driver, operator) has full control of the vehicle without intervention of the system. The system may still interact with other vehicles e.g., by providing status information. \\ \hline
Assisted operation & Human user (e.g., driver, operator) has control of the vehicle but may be assisted by active safety functions, and support/convenience functions. \\ \hline
Supervised \mbox{automation} & Human user monitors operation and intervenes for situations the vehicle is not designed to handle. The division of work may vary (e.g., system or human may perform lateral/longitudinal control, equipment control, performance of MRM) but supervision and some tasks are handled by a human user. 
%\review{text should discuss that tasks and handover must be 'fair' to both human and automation} 
\\ \hline
Unsupervised automation & No human user intervention is necessary, the vehicle is responsible for performing all operations, including MRM, safely. Human controls are optional (typically used for mixed-mode vehicles).  \\ \hline
\end{tabular}
\label{tab:av_automation}
\end{center}
\end{table}

\section{Conclusion}
\label{sec:conclusions}

This paper presents a unified taxonomy for AVs, including individual, cooperative, collaborative, on-road, and off-road vehicles. As mentioned, this first preprint version of the paper is released to allow for early feedback. In future revisions, we intend to expand the discussion to e.g.,  the implication on automation and interaction classes on terms such as ODD and DDT, further investigate the implications of automation of other tasks than the driving task, discuss architectural solutions, and implications for mixed traffic environments.

%Future work: Mixed traffic environments, dynamic operational task, ODDs.

\begin{comment}
\section*{Acknowledgement}
The preferred spelling of the word ``acknowledgment'' in American English is
without an ``e'' after the ``g.'' Use the singular heading even if you have
many acknowledgements. Avoid expressions such as ``One of us (S.B.A.) would
like to thank $\ldots$ .'' Instead, write ``F. A. Author thanks $\ldots$ .'' In most
cases, sponsor and financial support acknowledgements are placed in the
unnumbered footnote on the first page, not here.
\end{comment}

\bibliographystyle{unsrt}
\bibliography{refs}

\EOD

\end{document}